\documentclass[20pt,fleqn,leqno]{article}
\usepackage{graphicx,color,enumerate,amssymb}
\usepackage{cauchy}
\usepackage{amsmath}
\usepackage{amssymb}
\usepackage{ulem}
\usepackage{authblk}
\usepackage{natbib}

\renewcommand{\BBAA}{and}

\makeatletter
\@addtoreset{equation}{section}
\renewcommand\thetable{\@arabic\c@table}
\renewcommand\thefigure{\@arabic\c@figure}
\long\def\@makecaption#1#2{%
  \vskip\abovecaptionskip
  \begin{center}%
  \sbox\@tempboxa{#1: #2}%
  \ifdim \wd\@tempboxa >\hsize
    #1: #2\par
  \else
    \global \@minipagefalse
    \hb@xt@\hsize{\hfil\box\@tempboxa\hfil}%
  \fi
  \end{center}%
  \vskip\belowcaptionskip}
\def\N{{\rm I\kern-.15em N}}
\def\R{{\rm I\kern-.2em R}}
\def\Z{{\rm Z\kern-.26em Z}}
\makeatother

\newtheorem{thm}{Theorem}[section]

\newtheorem{prop}[thm]{Proposition}
\theorembodyfont{\rmfamily}
\newtheorem{rem}[thm]{Remark}

\newtheorem{ass}[thm]{Assumption}
\newcommand{\be}{\begin{eqnarray}}
\newcommand{\ee}{\end{eqnarray}}
\newcommand{\bq}{\begin{eqnarray*}}
\newcommand{\eq}{\end{eqnarray*}}

\newcommand{\stk}{\stackrel{\mbox{\scriptsize {\rm{P}}}}{\longrightarrow}}

\newcommand{\verk}{\stackrel{{\cal D}}{\longrightarrow}}

\newcommand{\bewend}{\hspace*{2mm}\rule{3mm}{3mm}}

\title{ Goodness--of--Fit Tests Based on the Min--Characteristic Function
}

\author[a,b]{Simos Meintanis\thanks{simosmei@econ.uoa.gr}}
\affil[a]{\small Department of Economics, National and Kapodistrian University
of Athens, Athens, Greece}
\author[c]{Bojana Milo\v{s}evi\'c\thanks{bojana.milosevic@matf.bg.ac.rs}}
\author[d]{ M.D. Jim\'enez--Gamero\thanks{dolores@us.es}}
\affil[b]{\small Pure and Applied Analytics\\ North--West University\\ Potchefstroom, South Africa}
\affil[c]{\small Faculty of Mathematics, University of Belgrade, Studenski trg 16, Belgrade, Serbia}
\affil[d]{\small Department of Statistics and Operations Research, University of Seville,
Seville, Spain}
\date{}

\begin{document}
\maketitle
\vspace*{1cm} {\small {\bf Abstract.} We propose tests of fit for classes of distributions that include the Weibull, the Pareto and the  Fr\'echet, distributions. The new tests employ the novel tool of the min--characteristic function and are based on an L2--type weighted distance between this function and its empirical counterpart applied on suitably standardized data. If data--standardization is performed  using the MLE of the distributional parameters then the method reduces to testing for the standard member of the family, with parameter values known and set equal to one.  We  investigate asymptotic properties of the tests, while a  Monte Carlo study is presented that includes the new procedure as well as competitors for the purpose of specification testing with three extreme value distributions.   
The new tests are also applied on a few real--data sets.}\\

\vspace*{0.3cm} {\small {\it Keywords.} min--characteristic function; Extreme--value distributions; Goodness--of--fit test; Invariant tests
\\ } 
\newpage

\section{Introduction}\label{sec_1}

\vspace*{-.3cm} The min--characteristic function (minCF) of a univariate random variable $X>0$ is defined as $\Psi_X(t)=\mathsf E(\min\{1,tX\}), \ t>0$. The minCF is a novel tool  introduced recently by \cite{falk2021min}, who show uniqueness, boundedness, and other interesting properties of $\Psi_X(\cdot)$. The same authors introduce and prove uniform strong consistency and weak convergence over compact intervals of the associated empirical minCF,
\be \label{ecf} 
\Psi_n(t)=\frac{1}{n} \sum_{j=1}^n \min\{1,tX_j\},
\ee
where $(X_j, \ j=1,...,n)$ denote independent copies of $X$.

Note in this connection that  for certain random variables, the minCF is often easier to compute than the classical characteristic function. This feature combined with the uniqueness of $\Psi_X(\cdot)$  for positively supported distributions, motivates  goodness--of--fit (GOF) tests based on some distance between the population minCF and the empirical minCF, $\Psi_n(\cdot)$. However, as with tests based on the characteristic function, the drawback with many such tests is with composite null hypotheses involving  parameters that actually occur in the population minCF of the distribution under test. These parameters are considered unknown and must be estimated from the data $(X_j, \ j=1,...,n)$ and it is often the case that the finite--sample as well as the asymptotic distribution of the resulting tests depends on the unknown true values of these parameters. If so, then GOF tests must be performed by bootstrap resampling methods that incur a certain computational cost on the implementation of the test procedure (see e.g. \cite{JG2009,JG2015}).

In this paper we suggest minCF--based GOF tests for a certain classes of composite null hypotheses that do not have the aforementioned computational drawback.  The paper is organized as follows. In Section \ref{sec_2} we introduce the null hypothesis to be tested and the corresponding  tests statistics. Then in Section   \ref{sec_3}  we prove that the resulting  tests are parameter--free if the distributional parameters are replaced by  maximum likelihood estimators (MLE), while in Section \ref{sec_4}  computational formulae are provided for the test statistic with specific instances of distributions under test. In Section \ref{sec_5} we present weak convergence of the test statistic under the null hypothesis and its behavior under alternatives. Section \ref{sec_6}  presents the results of a Monte Carlo study on the power of the new tests, as well as comparisons with other GOF tests. Real--data examples are included in Section \ref{sec_7}, and the paper concludes in Section \ref{sec_8} with discussion. All proofs are deferred to the Appendix.

\section{Goodness--of--fit tests}\label{sec_2}
Consider the null hypothesis  that the law of $X$ belongs to ${\cal{F}}$, for some $\vartheta \in \Theta$, 
where ${\cal{F}}=\{F_{\vartheta}; \vartheta \in \Theta\}$ is the particular class of distribution functions  indexed by a  parameter $\vartheta$ assumed to take values in  $\Theta$, the parameter space. 
The standard approach to GOF testing is by means of a distance measure between the population distribution function (DF) and its empirical counterpart, the empirical DF; a classic reference for empirical DF--based tests is \cite{dagostino1986tests}. Another approach which has become popular recently is the one based on the empirical characteristic function, while in the context of positive random variables the empirical Laplace transform has also attracted attention. The interested reader is referred to the papers of \cite{ndwandwe2023new}, \cite{ebner2022weibull}, \cite{cuparic2022new}, \cite{bothma2021characteristic}, \cite{lee2019inferential}, \cite{meintanis2016nonparametric}, \cite{milovsevic2016new}, \cite{meintanis2015goodness}, \cite{meintanis2003tests}, among others. 
\cite{meintanis2016review} gives a review of such GOF tests, while some earlier contributions for tests based on the characteristic function may by found in \cite[\S 3.9-3.10]{ushakov1999selected}.



We will introduce our test in $L^2(w)$, the (separable) Hilbert space setting  of measurable functions $f\in {\cal{H}}$,   $f: \mathbb R^+ \rightarrow \mathbb R$, which are squared integrable with respect to a 
weight function $w(\cdot)$. In this setting  the inner product of $f,g \in {\cal{H}}$ is defined as 
\[
<f,g>_w = \left(\int_0^\infty f(t)g(t)w(t)\:dt\right)^{1/2},
\]
with corresponding norm 
\[
\|f\|_w=(<f,f>)^{1/2} = \left(\int_0^\infty f^2(t)w(t)\:dt\right)^{1/2}.
\]

In complete analogy with the aforementioned tests based on the empirical DF (or the empirical characteristic function or the empirical Laplace transform), a reasonable test  procedure would be to reject the null hypothesis for large values of the test statistic     
\be \label{ecf_test}
T_{n,w}=\|\sqrt{n} \left(\Psi_n-\Psi_{\widehat \vartheta_n}\right)\|_w^2,
\ee
where $\Psi_\vartheta(\cdot)$ is the minCF corresponding to  $F_\vartheta$, 
$\Psi_n(\cdot)$ is the empirical minCF defined in \eqref{ecf}, and $\widehat \vartheta_n:=\widehat \vartheta_n(X_1,...,X_n)$ denotes some consistent estimator of $\vartheta$.   

Here however we will consider families of distributions with parameters $\vartheta=(c,\varphi)$, and such that the corresponding DF may be written as \be \label{fam} F_{c,\varphi}(x)=F_0\left(\left(\frac{x}{c}\right)^\varphi\right),\ee for some $c,\varphi>0$, and some fixed DF $F_0(\cdot)$. Clearly the corresponding family,   say ${\cal {F}}_0$, is completely determined by $F_0$, and then the null hypothesis may be restated as
\be \label{null}
\Upsilon_0: \mbox{ The law of $X$ $\in$ ${\cal{F}}_0$, for some $c, \varphi>0$.}
\ee
Specific instances of ${\cal{F}}_0$ to be considered in this work are the Weibull, the Pareto type I, and the  Fr\'echet   distributions, all belonging to the class of extreme--value distributions; see \cite{kotz2000extreme} for an account of extreme--value distributions.  


\section{Data transformations based on the MLE}\label{sec_3} 
In this section we investigate certain equivariance properties of the MLE under the class ${\cal{F}}_0$ with DF as in \eqref{fam}. These properties will have a direct impact on the proposed test, and specifically it will be shown that the test statistic will be rendered free of parameter values if the empirical minCF is applied on suitably standardized data.  To this end, let $(\widehat c_n,\widehat \varphi_n)$ be the MLE of $(c,\varphi)$, and consider the standardized observations  
\be \label{Y} \widehat Y_j=\left(\frac{X_j}{\widehat c_n}\right)^{\widehat\varphi_n}, \ j=1,...,n. \ee 

In the next theorem we investigate the aforementioned  equivariance properties of the MLE. For a general discussion of equivariant estimators we refer to \cite[\S29]{borovkov1998mathematical}, and \cite[Chapter 6]{romano2005testing}.

\begin{prop}
\label{prop} Let $(X_j, \ j=1,...,n)$, be independent copies on $X \in  {\cal{F}}_{0}$, for some ${\cal{ F}}_0$ and some $c,\varphi>0$. Assume further that the density of $X$ exists, and that the {\rm{MLEs}} $\widehat c_n(X_1,...,X_n)$ and $\widehat \varphi_n(X_1,...,X_n)$ of the parameters  $c$ and $\varphi$ also exist. Then the {\rm{MLE}} corresponding to $(\widehat Y_j, \ j=1,...,n)$ satisfy $\widehat c_n(\widehat Y_1,...,\widehat Y_n)=1$ and $\widehat \varphi_n(\widehat Y_1,...,\widehat Y_n)=1$.
\end{prop}

\medskip

An important consequence of Proposition \ref{prop} is that any GOF procedure for families satisfying \eqref{fam} that depends  on $(X_j, \ j=1,...,n)$ only via the observations $(\widehat Y_j , \ j=1,...,n)$ defined by \eqref{Y} 
satisfies 
\be \label{shapeinv} T_n(aX^{1/b}_1,...,aX^{1/b}_n)=T_n(X_1,...,X_n),\ee for each $a,b>0$, 
and consequently and without loss of generality, we may perform the test by assuming that   $c$ and $\varphi$ are fixed at $c=\varphi=1$. In this connection we suggest to apply the test defined by \eqref{ecf_test} by replacing the empirical minCF by
\be \label{mincf1}
\psi_n(t)=\frac{1}{n} \sum_{j=1}^n \min\{1,t\widehat Y_j\}
\ee
and the estimated population minCF $\Psi_{\widehat \vartheta_n}(\cdot):=\Psi_{\widehat c_n,\widehat \varphi_n}(\cdot)$ figuring in $T_{n,w}$ by $\psi_0(t):=\Psi_{1,1}(t)$. As a consequence, the finite--sample as well as the asymptotic distribution of the resulting test statistic, say ${\cal{T}}_{n,w}$, will be independent of the true but unknown values of $(c,\varphi)$.  On the practical level \eqref{shapeinv} implies that a potentially much simpler test may be invoked for testing families satisfying \eqref{fam}, such as in the case of the Weibull distribution whereby any test for exponentiality applied on $(\widehat Y_j,  \ j=1,...,n)$ may be used.

\section{Calculation of the test statistic}\label{sec_4}
In this section we particularize the families to be tested as per the null hypothesis in \eqref{null} and the functional form in $F_0(x)$ figuring in \eqref{fam}.  In this connection note that in the context of testing the null hypothesis $\Upsilon_0$ and in view of the results of the previous section,  we may perform the GOF test based  on the test statistic     
\be \label{testsum1}
{\cal {T}}_{n,w}:=\|\sqrt{n}\left(\psi_n-\psi_{0}\right)\|_w^2=n\left(\|\psi_n\|_w^2 +\|\psi_0\|_w^2-2 <\psi_n,\psi_0>_w\right),
\ee 
 for any given family  ${\cal{F}}_0$. 
In fact if we use $w(t)={\rm{e}}^{-\gamma t}, \ \gamma>0$, as weight function, the resulting test statistic, say ${\cal {T}}_{n,\gamma}$, may be written as  
\be \label{testsum2}
{\cal {T}}_{n,\gamma}=\frac{1}{n} \sum_{j,k=1}^n \Lambda_\gamma(\widehat Y_j,\widehat Y_k)+n {\cal{L}}_\gamma-2 \sum_{j=1}^n {\cal{\lambda}}_\gamma(\widehat Y_j) 
\ee
where
\[
\Lambda_\gamma(z_1,z_2)=\int_0^\infty \min\{1,t z_1\} \min\{1,tz_2\} {\rm{e}}^{-\gamma t} {\rm{d}}t,
\]
\[
{\cal{L}}_\gamma=\int_0^\infty \psi_0^2(t) {\rm{e}}^{-\gamma t} {\rm{d}}t,
\]
and
\[
\lambda_\gamma(z)=\int_0^\infty \min\{1,t z\} \psi_0(t) {\rm{e}}^{-\gamma t} {\rm{d}}t.
\]

The families to be consider are: \begin{itemize} 
\item the Weibull with $F_0(x)=1-{\rm{e}}^{-x}, \ x>0$, and $\psi_0(t)=t(1-{\rm{e^{-1/t}}})$,
\item the Pareto with $F_0(x)=1-x^{-1}, \ x>1$, and  $\psi_0(t)=t(1-\log t), \ t\leq 1$, and $\psi_0(t)=1, \ t>1$, 
\item the  Fr\'echet with $F_0(x)={\rm{e}}^{-x^{-1}}, \ x>0$, and $\psi_0(t)=1-{\rm{e}}^{-t}+t\Gamma[t]$, where $\Gamma[z]=\int_z^\infty u^{-1} {\rm{e}}^{-  u} {\rm{d}}u$. 
\end{itemize}

Clearly, the first integral $\Lambda_\gamma(z_1,z_2)$ does not depend on the family  ${\cal {F}}_0$ being tested and depends solely on the data $(\widehat Y_j, \ j=1,...,n)$ and the value of the weight parameter $\gamma$.  This integral is straightforward to compute. Specifically after some straightforward algebra we obtain 
\[
\Lambda_\gamma(z_1,z_2)= \frac{z_1}{z_2} \frac{2z^2_2-{\rm{e}}^{-\gamma/z_2}(2z^2_2+2\gamma z_2+\gamma^2)}{\gamma^3} +\frac{z_1}{z_2} \frac{{\rm{e}}^{-\gamma/z_2}(\gamma+z_2)}{\gamma^2}-z_1\frac{{\rm{e}}^{-\gamma/z_1}}{\gamma^2}, \ z_1\leq z_2.
\]
For $z_1> z_2$ we simply replace $z_1$ (resp. $z_2$) with $z_2$ (resp. $z_1$).  

On the other hand the second and third integral do depend on the family under test, with ${\cal{L}}_\gamma$ been independent of the observations whereas the third integral $\lambda_\gamma(\cdot)$, is the only one that involves both the family ${\cal {F}}_0$ as well as the observations. 

Specifically  for the Weibull distribution it follows after some long but otherwise routine calculations that the test statistic may be written as 
\be \label{weibulltest}
{\cal {T}}^{\rm W}_{n,\gamma}=\frac{1}{n} \sum_{j,k=1}^n \Lambda_\gamma(\widehat Y_j,\widehat Y_k)+n {\cal{L}}^{\rm W}_\gamma-2 \sum_{j=1}^n {\cal{\lambda}}^{\rm W}_\gamma(\widehat Y_j) 
\ee
where
\[{\cal{L}}^{\rm{W}}_\gamma=\frac{2}{\gamma^3}+\frac{4\sqrt{2}K_3[\sqrt{8 \gamma}]-4 K_3[\sqrt{4 \gamma}]}{\gamma^{3/2}},\]
\[
\lambda^{\rm W}_\gamma(z)=\frac{2z -{\rm{e}}^{-\gamma/z}(\gamma+2 z)}{\gamma^3}-(z M^{\rm W}_{\gamma,1}(z)+ M^{\rm W}_{\gamma,2}(z)),
\]
\[
M^{\rm W}_{\gamma,1}(z)=\int_0^{1/z} t^2 {\rm{e}}^{-(t^{-1}+\gamma t)}{\rm{d}}t
\]
and
\[
M^{\rm W}_{\gamma,2}(z)=\int_{1/z}^\infty t\: {\rm{e}}^{-(t^{-1}+\gamma t)}{\rm{d}}t,
\]
where $K_\nu[z]=(1/2)(z/2)^\nu \int_0^\infty {\rm{e}}^{-t-(z^2/4t)} t^{-(\nu+1)}{\rm{d}}t$,  stands for the modified Bessel function of the second kind of order $\nu$. 

By analogous calculations the test statistic for the Pareto distribution  may be written as
\be \label{paretotest}
{\cal {T}}^{\rm P}_{n,\gamma}=\frac{1}{n} \sum_{j,k=1}^n \Lambda_\gamma(\widehat Y_j,\widehat Y_k)+n {\cal{L}}^{\rm P}_\gamma-2 \sum_{j=1}^n {\cal{\lambda}}^{\rm P}_\gamma(\widehat Y_j) 
\ee
where
\[{\cal{L}}^{\rm{P}}_\gamma=\frac{{\rm{e}}^{-\gamma}}{\gamma}+\frac{{\rm{e}}^{-\gamma}(4-\gamma^2)+4(\log \gamma+\Gamma[\gamma]+\gamma_{\rm E}-1)}{\gamma^3}+I^{\rm P}_\gamma,\]
\begin{eqnarray*}
\lambda^{\rm P}_\gamma(z)&=&z\left(\frac{2-{\rm{e}}^{-\gamma}(2+2\gamma+\gamma^2)}{\gamma^3}-\frac{3-2 \gamma_{\rm E}-{\rm{e}}^{-\gamma}(3+\gamma)-2\Gamma[\gamma]-2\log \gamma}{\gamma^3}\right) \\ &+& \frac{z}{\gamma^2}({\rm{e}}^{-\gamma}(1+\gamma)-{\rm{e}}^{-\gamma/z}), \ z\leq 1,
\end{eqnarray*}
\begin{eqnarray*}
\lambda^{\rm P}_\gamma(z)&=&\frac{2z}{\gamma^3}-\frac{{\rm{e}}^{-\gamma/z}(2z^2+2\gamma z+\gamma^2)}{z\gamma^3}-z M^{\rm P}_{\gamma,1}(z) \\ &+&\frac{{\rm{e}}^{-\gamma/z}(\gamma+z)}{\gamma^2 z} -\frac{{\rm{e}}^{-\gamma}}{\gamma^2 }- M^{\rm P}_{\gamma,2}(z), \ z>1,
\end{eqnarray*}
\[I^{\rm P}_\gamma=\int_0^1 t^2 \log^2 t \: {\rm{e}}^{-\gamma 
t}{\rm{d}}t,\] 
\[
M^{\rm P}_{\gamma,1}(z)=\int_0^{1/z} t^2 \log t \: {\rm{e}}^{-\gamma t}{\rm{d}}t,
\]
and
\[
M^{\rm P}_{\gamma,2}(z)=\int_{1/z}^1 t \log t\: {\rm{e}}^{-\gamma t}{\rm{d}}t,
\] 
where $\gamma_{\rm E}(\approx 0.57721)$ denotes the Euler constant.

Finally for the   Fr\'echet distribution we have
\be \label{paretotest}
{\cal {T}}^{\rm F}_{n,\gamma}=\frac{1}{n} \sum_{j,k=1}^n \Lambda_\gamma(\widehat Y_j,\widehat Y_k)+n {\cal{L}}^{\rm F}_\gamma-2 \sum_{j=1}^n {\cal{\lambda}}^{\rm F}_\gamma(\widehat Y_j), 
\ee
where 
\[{\cal{L}}^{\rm{F}}_\gamma=\frac{1}{2+\gamma}+\frac{\gamma+2(\log(1+\gamma)+(1+\gamma)^{-1}-1)}{\gamma^2}-\frac{2(1+\gamma+\log(2+\gamma)+(2+\gamma)^{-1}-1)}{(1+\gamma)^2}+I^{\rm F}_\gamma,\]
\[
\lambda^{\rm F}_\gamma(z)=\frac{z(1-{\rm{e}}^{-\gamma/z})}{\gamma^2}-\frac{z(1-{\rm{e}}^{-(1+\gamma)/z})}{(1+\gamma)^2}+z M^{\rm F}_{\gamma,1}(z)+M^{\rm F}_{\gamma,2}(z),
\]
\[ I^{\rm F}_\gamma=\int_0^\infty t^2 \Gamma^2[t] \: {\rm{e}}^{-\gamma 
t}{\rm{d}}t, \]
\[
M^{\rm F}_{\gamma,1}(z)=\int_0^{1/z} t^2 \Gamma[t] {\rm{e}}^{-\gamma t}{\rm{d}}t,
\]
and
\[
M^{\rm F}_{\gamma,2}(z)=\int_{1/z}^\infty t \Gamma[t]\: {\rm{e}}^{-\gamma t}{\rm{d}}t.
\]

In the next section we investigate the asymptotic behaviour of the test ${\cal{T}}_{n,w}$ defined by \eqref{testsum1}, both under the null hypothesis 
$\Upsilon_0$ as well as under alternatives.  




\begin{rem}
 More general distributions with extra parameters may be included in our framework of testing GOF. For instance  the Burr type XII distribution belongs to the class of distributions defined by \eqref{fam}, with DF $F_0(x)=1-(1+x)^{-\xi}, \ \xi> 0$, and thus also satisfies Proposition \ref{prop}. Specifically if $X$ follows a Burr type XII distribution with parameters $(c,\varphi,\xi)$ then $aX^{1/b}$ follows a  Burr type XII distribution with parameters $(ac^{1/b},b\varphi,\xi)$. Other classes include members of the exponentiated class of distributions, like the exponentiated Weibull distribution with $F_0(x)=(1-{\rm{e}}^{-x})^\xi, \ \xi>0$ (see \cite{mudholkar1995exponentiated}), and the generalized gamma distribution with density  $f_{c,\varphi,\xi}(x)=(\varphi x^{\varphi \xi-1})/(c^{\varphi \xi}\Gamma(\xi)){\rm{e}}^{-(x/c)^\varphi}, \ \xi>0$.

\end{rem}

\section{Limit null distribution and consistency of the test}\label{sec_5}
\vspace*{-0.3cm} Here and in what
follows, the notation $\verk$ means convergence in distribution of
random elements and random variables, $\stk$ means  convergence in
probability,  $\stackrel{\mbox{\scriptsize {a.s.}}}{\longrightarrow} $ means almost sure convergence,
${\rm{o}}_{\mathsf{P}}(1)$ stands for convergence in probability to 0,
and ${\rm{O}}_{\mathsf{P}}(1)$ denotes a bounded in probability random variable or process. All limits are understood as $n\to \infty$, where $n$ stands for the sample size.

While for computational convenience we have suggested to actually perform the test based on the weight function $w(t)={\rm{e}}^{-\gamma t}, \ \gamma>0$, nevertheless the limit behaviour of the statistic  will be studied for arbitrary weight functions satisfying the following assumption.
\begin{ass} \label{weight}
 $w:[0,\infty) \mapsto (0,\infty) $,  is  continuous on $(0,\infty)$ and satisfies
$\int_0^\infty w(t)\:dt<\infty$.
\end{ass}


Recall also that the test figuring in \eqref{testsum1} implicitly involves the MLE via the transformed observations $(\widehat Y_j, \ j=1,...,n)$ defined in \eqref{Y}. In this connection we assume that $\widehat c_n$ and $\widehat \varphi_n$ admit the 
asymptotic representations in  next Assumption.

\begin{ass} \label{bahadur}
\be \label{bahc}
\sqrt{n}(\widehat c_n-c_0)=\frac{1}{\sqrt{n}}\sum_{j=1}^n {\rm {C}}(X_j)+\rm{o}_P(1)
\ee
and
\be \label{bahphi}
\sqrt{n}(\widehat \varphi_n-\varphi_0)=\frac{1}{\sqrt{n}}\sum_{j=1}^n \Phi(X_j)+\rm{o}_P(1), 
\ee
where $\mathsf E_0({\rm{C}}(X_1))=0=\mathsf E_0(\Phi(X_1))$, $\mathsf E_0({\rm{C}}^2(X_1))<\infty, \  \mathsf E_0(\Phi^2(X_1))<\infty$, and $\mathsf E_0(\cdot)$ indicating expectation taken under $F_{c_0,\varphi_0}$, with $(c_0,\varphi_0)$ being the true parameter values.
\end{ass}

\begin{rem}
The representations in Assumption \ref{bahadur} for the MLEs of $c$ and $\varphi$ may be derived by applying the technique in \S5.4.2 and \S6.2.1 of \cite{bickel2015mathematical}, whenever the family satisfies certain regularity conditions.  A family not meeting those conditions is the Pareto family. In this case, the MLE of the parameter $c$ is $X_{(1)}=\min\{X_1, \ldots, X_n\}$, but it may easily be checked that
\[
\sqrt{n}\left(X_{(1)}-c \right)=\rm{o}_P(1),
\]
that fulfills  \eqref{bahc} with ${\rm {C}}(x)=0, \forall x.$
\end{rem}
\bigskip
As already emphasized and due to test invariance with respect to $(c,\varphi)$, the study of the asymptotic null distribution of the test statistic ${\cal{T}}_{n,w}$ provided by Theorem \ref{thm_2_1} below, is carried out by setting $(c_0,\varphi_0)=(1,1)$.

It will be also assumed that $F_0$ has a density (with respect to the Lebesgue measure) as expressed in next assumption. 

\begin{ass} \label{density}
Let $a=\inf\{x>0: \, F_0(x)>0\}$, and  assume that $F_0$ has a density $f_0$ such that $f_0(a)=\displaystyle \lim_{x\to a^+}f_0(x)$ exists and is continuous on $[a,b]$, $\forall b>a$. Moreover suppose that 
\[
\int \log^\kappa(u)u^{\pm \rho}f_0^2(u)\: du<\infty, 
\]
for  
$\kappa=0$ and $\kappa=2$, and for some $0<\rho<1$.
\end{ass}

\begin{thm} \label{thm_2_1} 
Let $(X_j, \ j=1,...,n)$ be independent  copies of 
$X \in {\cal{F}}_{0}$. Suppose that Assumptions \ref{weight}, \ref{bahadur} and \ref{density} hold,  that $\mathsf E_0( \log^4(X))<\infty$ and that there exists $0<\varepsilon<1$ such that
\begin{equation} \label{w}
\int_1^\infty t^{8\varepsilon/(1-\varepsilon)} w(t)\: dt<\infty.   
\end{equation}
Then,
 \be \label{conT} 
 {\cal{T}}_{n,w} 
 \verk  \sum_{j \geq 1} \lambda_j \chi^2_{1j},
\ee
where $\chi^2_{11}, \chi^2_{12}, \ldots $ are independent chi-squared random variables each with one degree of freedom and $\{\lambda_j\}_{j \geq 1}$ are the eigenvalues of the operator
$A$ defined on $L^2(w)$ through
\[
Ag(x)=\int_0^{\infty}h_*(x,y)g(y)\: dy,
\]
with
\begin{eqnarray*}
h_*(x,y) & = & \int_{0}^{\infty} 
\left\{\min( 1,tx )-\psi_0(t)-\mu(t){\rm {C}}(x) \right\}
\left\{\min( 1,ty )-\psi_0(t)-\mu(t){\rm {C}}(y) \right\}
w(t) \: dt, \\
\mu(t) & = & -t\int_0^{1/t}u f_0(u)du. 
\end{eqnarray*}
\end{thm}

\begin{rem}
Notice that the weight function  $w(t)={\rm{e}}^{-\gamma t}$ adopted in Section \ref{sec_4}  satisfies Assumption \ref{weight} as well as  \eqref{w},  $\forall \, 0<\varepsilon<1$, for each $\gamma>0$.
\end{rem}

\begin{rem}
In Section \ref{sec_4}  we considered three special instances of distributions $F_0$. It can be easily checked that each of these distributions satisfies Assumption \ref{density}, $\forall \, \rho\in(0,1)$, and  that $\mathsf E_0( \log^4(X))<\infty$.
\end{rem}

\vspace*{.5cm} We now consider the asymptotic behavior of ${\cal{T}}_{n,w}$
under fixed alternatives to the null hypothesis $\Upsilon_0$. The following result implies
the consistency of the test that rejects $\Upsilon_0$ for
large values of ${\cal{T}}_{n,w}$.

\begin{thm} \label{cons} 
Let $(X_j, \ j=1,...,n)$ be independent copies of $X>0$. Assume that $(\widehat c_n,\widehat\varphi_n) \stk (c_X,\varphi_X)$, for some $c_X\in (0,\infty)$, $\varphi_X\in(0,\infty)$, and  that $\mathsf E( |\log(X)|)<\infty$. Then, 
\be \label{cons1}  
\frac{{\cal{T}}_{n,w}}{n} & \stk &\|\psi_X-\psi_0\|_w^2 :=\Delta_w,  \ee  \end{thm}
where $\psi_X(t)=\mathsf E(\min\{1,t(X/c_X)^{\varphi_X}\})$.  

\begin{rem}
If in the assumptions of Theorem  \ref{cons}, convergence in probability is replaced by almost sure convergence, i.e. if $(\widehat c_n,\widehat\varphi_n) \stackrel{\mbox{\scriptsize {a.s.}}}{\longrightarrow} (c_X,\varphi_X)$, then  we may write $\stackrel{\mbox{\scriptsize {a.s.}}}{\longrightarrow}$ in equation \eqref{cons1}.
\end{rem}


Under the standing assumptions, $\Delta_w$ is positive unless $\psi_X(t)$ is equal to 
$\psi_0(t)$ identically in $t$, i.e. if the minCF of $(X/c_X)^{\varphi_X}$ is equal to $\psi_0(\cdot)$. In turn by the uniqueness of the minCF this implies that $(X/c_X)^{\varphi_X}\in {\cal{F}}_0$, which,  since the transformation $X\mapsto (X/c_X)^{\varphi_X}$ is one--to--one,  entails  $X\sim F_{c,\varphi}$ with $(c,\varphi)=(c_X,\varphi_X)$. Thus   $\Delta_w>0$ unless $\Upsilon_0$ is true, and consequently the test that rejects the null hypothesis $\Upsilon_0$ for large values of ${\cal{T}}_{n,w}$ is consistent against each
fixed alternative distribution satisfying the assumptions stated
in Theorem \ref{cons}.

\section{Monte Carlo study}\label{sec_6}
In this section we study the finite--sample performance of the new GOF tests, simply denote it by $\cal T_\gamma$, in comparison with other tests. In order to perform comparisons, all tests are applied on standardized data \eqref{Y}  with sample size $n=20$ and $n=50$. In this connection and since the null distributions of tests statistics do not depend on unknown parameters, the tests' powers are obtained using a Monte Carlo procedure with $N=10,000$ replicates, with nominal level of significance $\alpha=0.05$. The study includes the following distributions: Weibull, Pareto, Gamma, Lognormal, Halfnormal, Linear failure rate, Fr\'echet, as well as Chen distributions, for some  specific choices of distributional parameters. These particular distributions occur  in analogous Monte Carlo studies;  see e.g. \cite{cuparic2020some}, \cite{allison2022distribution}, and \cite{ndwandwe2023new}. Their  density functions are given in Table \ref{tab:distributions}.

\begin{table}[]
    \centering
    \begin{tabular}{lll}\hline\hline
       Distribution  &Notation & Density function  \\\hline\hline
       Weibull  & $W(\varphi,c)$ & $\frac{\varphi x^{\varphi-1}}{c^\varphi} e^{-(\frac{x}{c})^\varepsilon},\;x>0;\;\varphi>0,c>0;$\\ 
       Pareto & $P(\varphi,c)$ &$\frac{\varphi c^{\varphi}}{x^{\varphi+1}}\;x>c;\;\varphi>0,c>0;$\\
       Gamma &$\Gamma(\varphi,c)$ & $\frac{x^{\varphi-1}}{c^{\varphi}\Gamma(\varphi)}e^{-\frac{x}{c}},\;x>0;\;\varphi>0,c>0;$\\
       Lognormal &$LN(\mu,\sigma)$ &$\frac{1}{x\sigma\sqrt{2\pi}}e^{-\frac{(\ln(x)-\mu)^2}{2\sigma^2}},\;x>0;\;\mu\in \mathbf{R},\sigma>0;$\\
       Halfnormal & $HN(\theta)$ & $\frac{\sqrt{2}}{\theta\sqrt{\pi}}e^{-\frac{x^2}{2\theta^2}},\;\; x>0;\;\theta>0$\\
       Linear faulure rate & $LFR(\theta)$ & $(1+\theta x)e^{- x- \frac{1}{2}\theta x^2},\;\; x>0;\;\theta>0;$\\
       Chen & $CH(\theta)$ & $2\theta x^{\theta-1}e^{x^{\theta}-2(1-e^{x^{\theta})}},\;x>0;\;\theta>0;$\\
       Fr\'echet & $F(\varphi,c)$ & $\frac{\varphi}{c}\big(\frac{x}{c}\big)^{-1-\varphi}e^{-(\frac{x}{c})^{-\varphi}},\;x>0;\;\varphi>0,c>0;$ 
       \\\hline
    \end{tabular}
    \caption{Distributions included in the power study}
    \label{tab:distributions}
\end{table}

For the Weibull distribution we include the Anderson--Darling test (often the best test amongst the classical tests based on the empirical DF), and the test of \cite{ozturk1988new}, which are labelled as AD and OK, respectively.
 We also include a recent test based on a Stein-type characterization. This test is denoted by $S_a$,  and we adopt the value $a=1$ for the value of tuning parameter following the authors' recommendation; see \cite{ebner2022weibull}. 
Another test which is related to the test based on  the minCF utilizes the max--characteristic function defined as $\Xi_X(t)=\mathsf E (\max\{1, tX\})$; see \cite{falk2017offspring}. The maxCF exists provided that $\mathsf E(X)<\infty$, and then the corresponding test, denoted by FS$_{\gamma}$, has a direct connection to a test based on the minCF, since $\Xi_X(t)=1+ t \mathsf E(X)- \Psi_X(t)$. Here we consider the maxCF test FS$_{\gamma}$ constructed analogously to our test with weight function $w(t)=e^{-\gamma t}$, with $\gamma=1$. We highlight that  tests based on the maxCF are also considered for the first time in the literature. 

The figures in Table \ref{tab:weibull} show percentage of rejection of the null hypothesis of a Weibull distribution for each test considered rounded to the nearest integer. Overall, the best performing test appears to be the OK test. For sample size $n=50$ our test with $\gamma=5$ is a good contender of OK, and preferable to the next best test, i.e. the AD test. For $n=20$,  our test ${\cal {T}}_5$ is less of a contender, but still ranks close to the AD test which is second best. In this connection we wish to point out that since the best performing  ${\cal {T}}_\gamma$--test across $\gamma$ is a formidable opponent of the OK test, data--driven selections of the weight parameter $\gamma$ such as the ones suggested by \cite{AS15} and \cite{Tenreiro},  seem to hold further promise for the new test. However more research is needed in this direction both in the actual implementation of such a test as well as for its asymptotics.

For the Pareto type I distribution we compare our test with  the $M_a$--test that is based on the difference between empirical characteristic function of transformed data and that of the standard uniform distribution (see \cite{meintanis2014class} and \cite{ndwandwe2023new} for details), and the test 
 based on the Mellin transform proposed by \cite{meintanis2009unified}, denoted by $G_a$, which is amongst the best performers in the Monte Carlo study of \cite{ndwandwe2023new}. Additionally, we include the likelihood ratio--based test $Z_C$ proposed by \cite{zhang2002powerful} and the test $I_2$ which is the one of the  best performers among the recent characterization--based tests utilizing the V-empirical distribution approach (see \cite{allison2022distribution}). The figures in Table \ref{tab:pareto} show clearly that the new tests based on $\cal T_\gamma$ are the most powerful nearly uniformly over alternatives and for all values of $\gamma$ with both sample sizes, followed by the $M_1$--based test.

Finally, for the Fr\'echet distribution  the proposed GOF test is compared against the likelihood ratio--based test $Z_A$ of \cite{zhang2002powerful} and the AD test. The figures shown in Table \ref{tab:frechet} show a similar picture to that of Table \ref{tab:weibull}, i.e., the new tests rank second best after the $Z_A$ test, being mostly more powerful than the AD--test, but not by a wide margin.

\begin{table}[h!]
    \centering
    \begin{tabular}{c|ccccccc|ccccccc|}
    \hline
     &\multicolumn{7}{c}{$n=20$}&\multicolumn{7}{c}{$n=50$}\\\hline
     Dist&$\mathcal{T}_{0.5}$&$\mathcal{T}_{1}$&$\mathcal{T}_{5}$  &AD&OK&$S_1$&FS$_1$ & $\mathcal{T}_{0.5}$&$\mathcal{T}_{1}$&$\mathcal{T}_{5}$  &AD&OK&$S_1$&FS$_1$ \\\hline
       
        $W(1,0.5)$&5&5&5&5&5& 5&4&5&6&5&5&5&5&5\\
         $W(0.5,1)$&5&5&5&6&5&5 &5&5&5&5&5&5&5&5\\\hline
         $\Gamma(0.8,1)$&5&5&5&6&5&6&5&5&6&6&6&5&6&6\\
           $\Gamma(2,1)$&6&6&7&6&7&5 &5&8&9&12&9&11&7&8\\
        $\Gamma(3,1)$&8&7&9&7&9&5 &7&13&13&18&13&13&10&12\\ 
        $LN(1)$ &23 &20 &26&23&30&17&20&56 &53&63&56&70&47&52  \\
        $LN(2.5)$ &24 &21 &26&22&30&18&19&57 &54&65&56&70&47&53 \\
        $HN(1)$ &7 &8 &8&9&6&10&7&11&13&16  &13&13&13&12\\
$LFR(0.2)$&55&47&43&55&56&39&47&95&92&89&97&96&94&92\\
        $LFR(0.5)$&49&41&39&48&52&32&41&92&88&85&94&93&85&87\\
         $LFR(0.8)$&45&37&37&43&47&31&38  &89&85&82&93&92&82&85\\
         $LFR(1)$&44&36&36&43&46&29&36  &89&83&81&91&91&80&84\\
         $CH(0.8)$&7&7&8&8&5&9  &7&11 &12 &16&12&13&11&12\\
         $CH(1)$&7&8&8&8&6&9  &7&11&12&15&12&12&11&12\\
         $CH(1.5)$&8&8&8&8&6&9  &7&11&12&16&12&12&11&12\\
         \hline
    
    \end{tabular}
    \caption{GOF tests for the Weibull distribution: Empirical percentage of rejection}
    \label{tab:weibull}
\end{table}


\begin{table}[h!]
    \centering
    \begin{tabular}{c|ccccccc|ccccccc|}
    \hline
     &\multicolumn{7}{c}{$n=20$}&\multicolumn{7}{c}{$n=50$}\\\hline
        Dist&$\mathcal{T}_{0.5}$&$\mathcal{T}_{1}$&$\mathcal{T}_{5}$  &$M_1$&$G_2$&$Z_C$&$I_2$& $\mathcal{T}_{0.5}$&$\mathcal{T}_{1}$&$\mathcal{T}_{5}$  &$M_1$&$G_2$&$Z_C$&$I_2$\\\hline
       
        ${P}(1,1)$&5&5&5&5&5&5
        &4&5&5&5&5&5&5&5\\
         ${P}(2,1)$&5&5&5&5&5&5 &4&5&5&5&5&5&5&5\\\hline
          $W(0.5,1)+1$&29 &27 &19 &34&34 &51&39&59&57&43&66&89&89&79\\
          $W(0.8,1)+1$&6 &5 &7 &3&2 &6&4&16&17&24&12&7&13&6\\
         $W(1.2,1)+1$&38 &36 &38 &27&17 &5&22&94&94&95&91&88&61&80\\  
          $W(1.5,1)+1$&62 &61 &63 &53&38 &12&42&100&100&100&99&98&93&97\\  
           $\Gamma(0.8,1)+1$&7&7&8&4&2 &6&4&22&23&29&16&10&12&8\\
         $\Gamma(1,1)+1$&18&18&19&12&7 &4&9&66&66&70&59&51&22&40\\
         $\Gamma(1.2,1)+1$&34&32&34&24&15 &4&18&91&91&93&88&83&53&76\\
          $HN(1)+1$&47&51&55&37&25&9&29&98&98&99&95&92&75&87\\
          $LN(1)+1$&15 &14 &13 &11&7 &2&9&63&64&56&65&61&23&55\\  
         $LN(1.2)+1$&7 &7 &6 &5&3 &2&5&32&31&28&32&29&7&24\\ 
          $LN(1.5)+1$&4 &4 &4 &3&2 &3&4&6&6&7&6&4&3&4\\
           $LN(2.5)+1$&24 &26 &18 &31&31 &34&39&55&52&38&64&65&66&72\\ 
          $LFR(0.2)+1$&12 &12 &14 &8&4 &3&5&50&48&53&4&35 &13&27\\  
          $LFR(0.5)+1$&18 &18 &19&13 &6&4&5&65&65&69&58&49&20&41\\ 
           $LFR(0.8)+1$&22 &22 &23 &15&8 &4&8&76&75&79&68&59&27&50 \\
            $LFR(1)+1$&24 &23 &25 &17&4 &9&11&79&80&81&62&64&32&54 \\
            $CH(0.8)+1$&6 &6 &8 &4&2 &8&3&19&19&26&13&5&13&10 \\
            $CH(1)+1$&22 &22 &25 &16&4 &5&7&77&78&82&67&59&33&48 \\
             $CH(1.5)+1$&73 &74 &76 &64&47 &19&50&100&100&100&99&99&98&99 \\
            \hline
                \end{tabular}
    \caption{GOF tests for the Pareto distribution: Empirical percentage of rejection}
    \label{tab:pareto}
\end{table}

\begin{table}[h!]
    \centering
    \begin{tabular}{c|ccccc|ccccc|}
    \hline 
    &\multicolumn{5}{c}{$n=20$}&\multicolumn{5}{c}{$n=50$}\\\hline
        Dist&$\mathcal{T}_{0.5}$&$\mathcal{T}_{1}$&$\mathcal{T}_{5}$ &AD &$Z_A$& $\mathcal{T}_{0.5}$&$\mathcal{T}_{1}$&$\mathcal{T}_{5}$ &AD  &$Z_A$ \\\hline  
       ${F}(2,1)$  & 5& 5& 5& 5&5&5&5&5&5&5\\
       ${F}(1,0.5)$ & 5& 5&5& 5&5&5&5&5&4&5\\\hline
       $W(0.8,1)$ &72 &73 &73&69&79&99&99&99&98&82\\
       $W(1.5,1)$ &72 &72 &73&68&80&99&99&99&99&99\\
        $\Gamma(0.8,1)$ &76 &77 &78&78 &82&99&99&100&99&99\\
       $\Gamma(1,1)$ &72&72 &73&74&78&99&99&99&99&99\\
       $\Gamma(2,1)$ &60 &58 &58&54&66&96&96&97&95&98 \\
           $\Gamma(5,1)$ &48 &46&46&43&46&88&89&89&87&94 \\
       $LFR(0.5)$ &7 &7 &8&9&5 &21&18&23&22&23 \\
       $LFR(0.8)$ &8 &7 &8&8&5&22&19&23&22&23 \\
       $LFR(1)$ & 7& 7&9&8&5&20&18&24&22&24\\
       $LN(0.8)$ &25 &24 &22&20&29&60 &60 &58&54&69\\
       $LN(1.5)$ &26 &23 &23&21&28&61&69 &59 &54&70\\
       $HN(1)$ &82 &83 &86&81&89&99&99 &100 &99&100\\
       \hline
    \end{tabular}
    \caption{GOF tests for the Fr\'echet distribution: Empirical percentage of rejection}
    \label{tab:frechet}
\end{table}



\section{Real--data applications}\label{sec_7}

The Weibull distribution is often used  in the context of reliability analysis.  In this connection, \cite{wu2021parameter}  consider the Weibull distribution for strength--modeling of two glass fibers, of length 1.5cm and 15 cm, respectively. (The datasets are originally taken from  \cite{smith1987comparison}). The main goal of their study was to consider different methods for parameter estimation, while the fit with Weibull distribution has been justified using the classical Kolmogorov-Smirnov test. However, it is not clear how this test was used, i.e. whether the fact that it is not distribution free in the presence of estimation of unknown parameters was taken into account. Here we apply the tests utilized in our power study to check goodness--of--fit to the Weibull distribution. The results are presented in Table \ref{tab:realdata}. The p--values reported suggest   that  for Dataset 2 the Weibull distribution may be a reasonable model supported by all tests considered. On the other hand, the fit of Dataset 1 to the Weibull distribution is questionable, with the exception of the OK test and marginally so by the new test for $\gamma=5$.

Next we turn to the fit to the  Fr\'echet distribution.  In this connection, \cite{ramos2020frechet} present an overview of the properties of this distribution and, as an illustration, they considered  five real data sets related to minimum monthly flows of water (m3/s) in the Piracicaba River, located in S\~{a}o Paulo state, Brazil. They concluded,  using several information criteria, that in all cases the  Fr\'echet distribution is the best of the models considered. The p--values reported in Table \ref{tab:realdata2} seem to corroborate the conclusions of \cite{ramos2020frechet} by means of the formal GOF tests applied herein.  

As a final illustration, we apply our tests  to the monetary expenses incurred as a result of wind related catastrophes in 40 separate instances during 1977 (after the de-grouping algorithm used in \cite{allison2022distribution}). The results presented in Table \ref{tab:realdata3} indicate that the Pareto distribution may be one of the possible choices for modeling these data.

\begin{table}[]
    \centering
   \begin{tabular}{|cc|ccccccc|}
    \hline
     &&\multicolumn{7}{|c|}{Test statistics}\\\hline
&$n$&$\mathcal{T}_{0.5}$&$\mathcal{T}_{1}$&$\mathcal{T}_{5}$  &AD&OK&$S_1$&FS$_1$   \\\hline
 Dataset 1 &63& 0.0066& 0.0019& 0.0693&0.0021&0.2676&0.0021&0.019\\
 Dataset 2 &46&0.5508&0.3566&0.1666&0.526&0.1268& 0.2649 &0.3602\\\hline
    \end{tabular}
    \caption{Glass fiber strength: p-values for tests for the Weibull distribution  }
    \label{tab:realdata}
\end{table}

\begin{table}[]
    \centering
   \begin{tabular}{|cc|ccccc|}
    \hline
     &&\multicolumn{5}{|c|}{Test statistics}\\\hline
&$n$&$\mathcal{T}_{0.5}$&$\mathcal{T}_{1}$&$\mathcal{T}_{5}$  &AD& $Z_A$  \\\hline
 May  &40&0.996&0.9945&0.9607&0.9955&0.9780\\
 June&39&0.2533&0.3376&0.5746&0.1163&0.1961\\
 July&39&0.0666&0.0618&0.1057&0.1139&0.2947\\
August&41&0.3216&0.2968&0.1981&0.4738&0.5863\\
September&41&0.3216&0.2968&0.1981&0.4768&0.5854\\\hline
    \end{tabular}
    \caption{Minimum monthly flows of
water: p-values for tests for the Frech\'et distribution }
    \label{tab:realdata2}
\end{table}

\begin{table}[]
    \centering
   \begin{tabular}{|c|ccccccc|}
    \hline
     &\multicolumn{7}{|c|}{Test statistics}\\\hline
$n$&$\mathcal{T}_{0.5}$&$\mathcal{T}_{1}$&$\mathcal{T}_{5}$ & $M_1$&$G_2$&$Z_C$&$I_2$\\\hline
 40 &0.1958 &0.1916 &0.1036&0.2926&0.2794&0.2686 &0.9376\\
 \hline
    \end{tabular}
    \caption{Wind catastrophes: p-values for tests for the Pareto distribution }
    \label{tab:realdata3}
\end{table}

\section{Conclusion and outlook}\label{sec_8}
We propose for the first time GOF tests based on the minCF. The tests are computationally convenient and may be applied to arbitrary positively supported distributions with unknown parameters. It is shown that for certain families of such laws the implementation of the new tests on the basis of MLE--standardized data lead to simplified procedures that mimic simple, rather than composite, GOF tests with fixed parameter values. The asymptotic null distribution of the test statistic is obtained under general conditions, and moreover it was shown that the suggested tests are consistent against arbitrary deviations from the null hypothesis. A Monte Carlo study is conducted in order to assess the performance of the new tests against competitors for the cases of Weibull, Pareto type I and  Fr\'echet, distributions. Our findings show that the minCF--based  tests could in certain cases be the best option, but most importantly, and in view of the fact that there can not be any ``true" best test that outperforms competitors uniformly over all alternatives (see \cite{Janssen} and \cite{Escanciano}), that even in the cases that the new tests are not most powerful, they often rank near the top. We conclude with a few real--data applications that complement earlier findings.                       

\section{Appendix}
\noindent {\sc Proof of Proposition 3.1:} We will show that the MLE of $(c,\varphi)$ based on $Y_j=a X_j^{1/b}, \ j=1,...,n$, satisfy  
\be \label{c} \widehat c_n(Y_1,...,Y_n)=a\:\widehat c^{1/b}_n(X_1,...,X_n),\ee  
and 
\be \label{phi}  \widehat \varphi_n(Y_1,...,Y_n)=b\:\widehat \varphi_n(X_1,...,X_n),\ee  
for each $a,b>0$, and then the result trivially holds by applying \eqref{c} and \eqref{phi} on $(\widehat Y_j, \ j=1,...,n)$, with $(a,b)$ set equal to $(\widehat c_n^{-\widehat \varphi_n}, \widehat \varphi^{-1}_n)$.   

In order to proceed with the proof we first compute  
from \eqref{fam} the density corresponding to $F_{c,\varphi}$ as    
\be \label{den}
f_{c,\varphi}(x)=\frac{{\rm{d}}F_{c,\varphi}(x)}{{\rm{d}}x}
=\frac{{\rm{d}}F_0((\frac{x}{c})^\varphi)}{{\rm{d}}x}=\frac{\varphi}{c}
 \left(\frac{x}{c}\right)^{\varphi-1}f_0\left(\left(\frac{x}{c}\right)^\varphi\right),
\ee where $f_0(x):=f_{1,1}(x)=\frac{{\rm{d}}F_0(x)}{{\rm{d}}x}$.  
Also note that if $X\sim F_{c,\varphi}$, then $a X^{1/b} \sim F_{\sigma,\kappa}$, where $(\sigma,\kappa)=(ac^{1/b},b \varphi)$ for each $a,b>0$. 

Then we calculate the likelihood function corresponding to the transformed parameters $(\sigma,\kappa)$ and the transformed observations $(Y_j, \ j=1,...,n)$  as
\be \label{mle2}
{\cal{L}}(Y_1,...,Y_n;\sigma,\kappa)=\frac{\kappa^n}{\prod_{j=1}^n Y_j} 
  \prod_{j=1}^n \left(\frac{Y_j}{\sigma}\right)^\kappa  f_0\left(\left(\frac{Y_j}{\sigma}\right)^\kappa\right).
  \ee
After some further algebra we obtain  
\be \label{mle3} {\cal{L}}(Y_1,...,Y_n;\sigma,\kappa)=D(X_1,...,X_n;a,b){\cal{L}}(X_1,...,X_n;c,\varphi)
\ee
where 
\[ {\cal{L}}(X_1,..., X_n;c,\varphi)=\frac{\varphi^n}{\prod_{j=1}^n X_j} 
\prod_{j=1}^n \left(\frac{X_j}{c}\right)^\varphi  f_0\left(\left(\frac{X_j}{c}\right)^\varphi\right), 
\] is the likelihood function corresponding to the original  parameters $(c,\varphi)$ and the original observations $(X_j,\ j=1,...,n)$, and   
\[
D(X_1,...,X_n;a,b)= \left(\frac{b}{a}\right)^n  \prod_{j=1}^n X^{1-(1/b)}_j. 
\]
Equation \eqref{mle3} implies  that the two likelihoods are equivalent within a proportionality constant and thus  are both maximized by setting $(c,\varphi)$ equal to the  MLE 
$(\widehat c_n(X_1,...,X_n),\widehat \varphi_n(X_1,...,X_n))$. 
In order to complete the proof we should show that 
\[ \label{mle4}
{\cal{L}}(Y_1,...,Y_n;\widehat \sigma_n,\widehat \kappa_n)=D(X_1,...,X_n;a,b){\cal{L}}(X_1,...,X_n;\widehat c_n,\widehat \varphi_n),
\]
where $(\widehat \sigma_n,\widehat \kappa_n)=(a \widehat c^{1/b}_n(X_1,...,X_n) ,b\widehat \varphi_n(X_1,...,X_n))$. This is done by  replacing in \eqref{mle2}, $(\sigma,\kappa)$ by  $(\widehat \sigma_n,\widehat \kappa_n)$ and $Y_j$ by $a X^{1/b}_j, \ j=1,...,n$,  and by performing some straightforward algebra. For a different proof of this proposition we refer to \cite{matsui2023mle}. \bewend

\bigskip

\noindent {\sc Proof of Theorem \ref{thm_2_1}:}  Notice that 
\[
{\cal{T}}_{n,w}=\frac{1}{n}\sum_{j,k=1}^n h(X_j,X_k; \widehat c_n, \widehat \varphi_n), 
\]
with
\[
h(X_j,X_k; c, \varphi)=\int_0^\infty g(t,X_j; c, \varphi)g(t,X_k; c, \varphi)w(t) \: dt,
\]
\[
g(t,x; c, \varphi) = \min \{ 1,t(x/c)^\varphi \}-\psi_0(t),
\]
that is,  ${\cal{T}}_{n,w}$ is $n$ times a degree 2 $V$-statistic, where certain parameters have been replaced with estimators. A useful result to  derive the asymptotic null distribution of these type of statistics is Theorem 2.16 in de Wet and Randles (1987). The proof will check that the conditions in such theorem hold true in our setting, implying that the claimed result is true. Specifically, we will check that Condition 2.9, Condition 2.10, Condition 2.11 and display (2.17) in de Wet and Randles (1987) are met.

\medskip

\noindent \underline{Condition 2.9 in de Wet and Randles (1987)} Recall that $\mathsf E_0(\cdot)$ stands for the expectation under the null hypothesis with $(c,\varphi)=(1,1)$. Let
$$ \mu(t; c, \varphi)= \mathsf E_0[g(t,X; c, \varphi)].$$
Since 
\begin{equation} \label{g.bounded}
|g(t,X; c, \varphi)| \leq 2, 
\end{equation}
$ \mu(t; c, \varphi)$ exits 
$\forall c, \varphi, t>0$  and satisfies $ \mu(t; 1, 1)=0$. We must show that
$\mu(t; c, \varphi)$ has an $L_2(w)$ differential at $(c, \varphi)=(1,1)$, call it $d_1\mu(t; 1,1)=(d_{1c}\mu(t; 1,1), d_{1 \varphi}\mu(t; 1,1))$, with $d_{1c}\mu(t; 1,1)$, $d_{1 \varphi}\mu(t; 1,1)$ $\in L^2(w)$. The existence of an  $L_2(w)$ differential  means that for any $\eta>0$, there is a ball centred at $(c,\varphi)=(1,1)$, say $B$, such that $(c,\varphi) \in B$ implies that
\begin{equation} \label{diff}
\| (c,\varphi)-(1,1)\|^{-2}\int_0^\infty \left\{\mu(t; c, \varphi)-d_{1c}\mu(t; 1,1)(c-1)-d_{1 \varphi}\mu(t; 1,1)(\varphi-1) \right\}^2w(t)\:dt<\eta.
\end{equation}
We have that,
\begin{eqnarray*}
 \mu(t; c, \varphi) & =  & \frac{t}{c^\varphi}\int_0^{c(1/t)^{1/\varphi}}x^\varphi f_0(x)\: dx+1-F_0(c(1/t)^{1/\varphi})-\psi_0(t)\\
 & := & I(t; c, \varphi) +1-F_0(c(1/t)^{1/\varphi})-\psi_0(t),\\
\frac{\partial}{\partial c}\mu(t; c, \varphi) & =  &-\frac{\varphi}{c}I(t; c, \varphi),\\
\frac{\partial}{\partial \varphi}\mu(t; c, \varphi) & =  & -\log(c) I(t; c, \varphi),\\
\frac{\partial^2}{\partial c^2}\mu(t; c, \varphi) & = & \frac{\varphi(\varphi+1)}{c^2} I(t; c, \varphi)  -\frac{\varphi}{c}\left( 1/t\right)^{1/\varphi}f_0\left(c \left(1/t\right)^{1/\varphi}\right),\\
\frac{\partial^2}{\partial c \partial \varphi}\mu(t; c, \varphi) & =  & \frac{\varphi \log(c)-1}{c} I(t; c, \varphi)-\frac{\log(t)}{\varphi} \left( 1/t\right)^{1/\varphi}f_0\left(c \left(1/t\right)^{1/\varphi}\right),\\
\frac{\partial^2}{\partial \varphi^2}\mu(t; c, \varphi) & =  & \log^2(c)
I(t; c, \varphi)-c\frac{\log(c)}{\varphi} \log(t) \left( 1/t\right)^{1/\varphi}f_0\left(c \left(1/t\right)^{1/\varphi}\right).
\end{eqnarray*}
 Now,
\begin{eqnarray}
 d_{1c}\mu(t; 1,1) & = &  \left. \frac{\partial}{\partial c}\mu(t; c, \varphi)  \right|_{(c, \varphi)=(1,1)}=-I(t;1,1):=\mu(t), \label{deriv1}\\ 
 d_{1 \varphi}\mu(t; 1,1) & = & \left. \frac{\partial}{\partial \varphi}\mu(t; c, \varphi)  \right|_{(c, \varphi)=(1,1)}=0. \nonumber 
\end{eqnarray}
As 
\begin{equation}\label{I}
0 \leq I(t; c, \varphi)  \leq 1,  \quad \forall c, \varphi, t>0,
\end{equation}
it readily follows that $d_{1c}\mu(t; 1,1), d_{1 \varphi}\mu(t; 1,1) \in L^2(w)$.

Let $H(t; c, \varphi)$ be the $2 \times 2$ matrix of second order partial derivatives of $\mu(t; c, \varphi)$ with respect to $\mu$ and $\varphi$, and   let $\|\cdot \|_F$ denote the Frobenius norm. Taking into account \eqref{I}, 
and Assumption \ref{density}, routine calculations show that  for any $(c, \varphi)$ such that $\| (c,\varphi)-(1,1)\|\leq \rho$, 
\begin{equation}\label{I_H}
I_H(c,\varphi)=\int \|H(t; c, \varphi)\|_F^2w(t)\: dt \leq M,
\end{equation}
for certain positive constant $M$. 

By Taylor expansion, 
\[ 
\mu(t; c, \varphi)-\mu(t; 1,1)-d_{1c}\mu(t; 1,1)(c-1)=0.5
(c-1, \varphi-1) H(t; \widetilde{c}, \widetilde{\varphi})(c-1, \varphi-1)^\top
\] 
where $(\widetilde{c}, \widetilde{\varphi})$ is between $(c,\varphi)$ and $(1,1)$,
for each $t>0$, and hence
\begin{equation}\label{Te}
\left\{\mu(t; c, \varphi)-\mu(t; 1,1)-d_{1c}\mu(t; 1,1)(c-1) \right\}^2 \leq 0.25
\|(c-1, \varphi-1)\|^4 \|H(t; \widetilde{c}, \widetilde{\varphi})\|_F^2.
\end{equation}
From  \eqref{I_H} and \eqref{Te}, it readily follows that   \eqref{diff} holds true for $B$ the ball centered at $(c, \varphi)=(1,1)$ with radius $\rho$.

\medskip

\noindent \underline{Condition 2.10 in de Wet and Randles (1987)}. This is our Assumption \ref{bahadur}.

\medskip

\noindent \underline{Condition 2.11 (a) in de Wet and Randles (1987)}. We must show that there is a number $M>0$ and a neighborhood $\cal N$ of $(c,\varphi)=(1,1)$ such that if $(c,\varphi)\in {\cal N}$ and the open ball centered at $(c,\varphi)$ with radius $\delta$, $B((c,\varphi),\delta)$, is such that $B((c,\varphi),\delta)\subset {\cal N}$, then
\begin{equation} \label{cond211a}
\int_0^\infty \mathsf E^2\left[ \sup_{(c_1,\varphi_1) \in B((c,\varphi),\delta)} \left| g(t,X; c_1,  \varphi_1)-g(t,X; c, \varphi)\right|\right]w(t)\: dt \leq M \delta^2.
\end{equation}
We have that
\[
\left| g(t,x; c_1,  \varphi_1)- g(t,x; c, \varphi)\right| \leq t 
\left| \left(\frac{x}{c_1} \right)^{\varphi_1}-\left(\frac{x}{c} \right)^{\varphi} \right|, \]
if  $0<x<\max \left\{c \left(1/t \right)^{1/\varphi}, c_1 \left(1/t \right)^{1/\varphi_1} \right\}$, and $g(t,x; c_1,  \varphi_1)- g(t,x; c, \varphi)=0$ otherwise.
Taking into account that
\[  
\frac{\partial}{\partial c} \left(\frac{x}{c} \right)^{\varphi}  =  -\frac{\varphi}{c}  \left(\frac{x}{c} \right)^{\varphi} , 
\quad
\frac{\partial}{\partial \varphi} \left(\frac{x}{c} \right)^{\varphi} =  \log\left(\frac{x}{c}\right)  \left(\frac{x}{c} \right)^{\varphi} ,
\]
applying the mean value theorem and the  Cauchy–Schwarz inequality, one gets that if $(c_1,\varphi_1) \in B((c,\varphi),\delta)$ then
\begin{equation} \label{cota1}
\left| g(t,x; c_1,  \varphi_1)- g(t,x; c, \varphi)\right| \leq \delta t 
\left( \frac{x}{\widetilde c}\right)^{\widetilde \varphi}
\left \| \left(-\frac{\widetilde \varphi}{\widetilde c}, 
\log\left( \frac{x}{\widetilde c}\right)\right) \right\|,
\end{equation}
 $0<x<\max \left\{c \left(1/t \right)^{1/\varphi}, c_1 \left(1/t \right)^{1/\varphi_1} \right\}$, where $(\widetilde c, \widetilde \varphi)=\alpha (c,\varphi)+(1-\alpha)(c_1, \varphi_1),$ for some $0<\alpha<1$ (the value of $\alpha$ may depend on $t$ and on $x$). Let  ${\cal N}=B((1,1),\varepsilon/2)$, which implies that $\delta< \varepsilon/2$,
 \[
 1-\varepsilon< c,\,c_1,\, \widetilde{c} <1+\varepsilon, \quad
 1-\varepsilon< \varphi, \, \varphi_1, \,  \widetilde{\varphi} <1+\varepsilon,
 \]
 and
\[
\frac{1-\varepsilon}{1+\varepsilon} \leq \frac{ \widetilde{\varphi}}{\varphi}=
\alpha+(1-\alpha)\frac{\varphi_1}{\varphi}, \,  \frac{ \widetilde{\varphi}}{\varphi_1}=
1-\alpha+ \alpha\frac{\varphi}{\varphi_1}\leq \frac{1+\varepsilon}{1-\varepsilon}.
\]
Thus,
\begin{equation} \label{cota2}
0<t x^{\widetilde \varphi}< t \max \left\{c \left(1/t \right)^{1/\varphi}, c_1 \left(1/t \right)^{1/\varphi_1} \right\}^{\widetilde \varphi} \leq
(1+\varepsilon)^{1+\varepsilon}  \max \left\{ t^{2\varepsilon/(1-\varepsilon)}, 
 t^{2 \varepsilon/(1+\varepsilon)} \right\}.
    \end{equation}
From \eqref{cota1} and  \eqref{cota2},  
\begin{equation} \label{cota}
\begin{array}{c}
\displaystyle \left| g(t,x; c_1,  \varphi_1)- g(t,x; c, \varphi)\right|^2   \leq \\  \displaystyle \delta^2 \left(\frac{1+\varepsilon}{1-\varepsilon}\right)^{2(1+\varepsilon)} \left\{ \left(\frac{1+\varepsilon}{1-\varepsilon}\right)^2+2\log^2(x)+2\log^2(1+\varepsilon) \right\}
\max \left\{ t^{4 \varepsilon/(1-\varepsilon)}, 
 t^{4\varepsilon/(1+\varepsilon)} \right\} .
 \end{array}
\end{equation}
Notice that
\begin{equation*}  
I=\int_{0}^{\infty} \max \left\{ t^{4 \varepsilon/(1-\varepsilon)}, 
 t^{4 \varepsilon/(1+\varepsilon)} \right\} w(t) \: dt \leq \int_0^1 t^{4 \varepsilon/(1+\varepsilon)}  w(t) \: dt+\int_{0}^{\infty}  t^{4 \varepsilon/(1-\varepsilon)} w(t) \: dt:=I_1+I_2.
\end{equation*}
Clearly $0<I_1<\infty$ as it is the integral of a continuous function on a bounded interval; from \eqref{w} we also have that $0<I_2<\infty$.
Since $0<I<\infty$ and  $\mathsf E_0( \log^4(X))<\infty$, from \eqref{cota}   it readily follows that \eqref{cond211a} holds true.

\medskip

\noindent \underline{Condition 2.11 (b) in de Wet and Randles (1987)}  We must show that  for any $\eta>0$, there is a $\delta^*$ such that $0<\delta<\delta^*$, $(c,\varphi) \in {\cal N}$ and $B((c,\varphi),\delta)\subset {\cal N}$ imply that
\begin{equation} \label{cond211b}
\int_0^\infty \mathsf E\left[ \sup_{(c_1,\varphi_1) \in B((c,\varphi),\delta)} \left| g(t,X; c_1,  \varphi_1)-g(t,X; c, \varphi)\right|^4\right]w(t)\: dt \leq \eta.
\end{equation}
Proceeding as in the proof of Condition 2.11 (a), it can be seen that $ \int_{0}^{\infty} \max \left\{ t^{8 \varepsilon/(1-\varepsilon)}, 
 t^{8 \varepsilon/(1+\varepsilon)} \right\}$ $w(t) \: dt<\infty$.
Taking  $\delta^*=\varepsilon/2$, from  \eqref{cota}   and   $\mathsf E_0( \log^4(X))<\infty$ it readily follows that \eqref{cond211b} holds true.

\medskip

\noindent \underline{(2.17) in \cite{de1987effect}}
We must check that 
\begin{equation} \label{cond217}
\mathsf E_0[h_*^2(X_1,X_2)]<\infty, \quad \mathsf E_0[|h_*(X_1,X_1)|]<\infty. 
\end{equation}
With this end, notice that
\[
h_*(x,y)=\int_0^\infty \{g(x,t; 1,1)-\mu(t){\rm {C}}(x)\} \{g(y,t; 1,1)-\mu(t){\rm {C}}(y)\}w(t)\:dt.
\]
Now, using \eqref{g.bounded}, \eqref{deriv1} and\eqref{I}, we get that
\begin{equation} \label{h.bound}
|h_*(x,y)| \leq M_1+M_2{\rm {C}}(x)+M_3{\rm {C}}(y)+M_4{\rm {C}}(x){\rm {C}}(y),
\end{equation}
for certain positive constants $M_1,$ $M_2,$ $M_3,$ $M_4$. Finally, from Assumption \ref{bahadur} and \eqref{h.bound}, it is concluded that  \eqref{cond217} holds true \bewend

\bigskip

\noindent {\sc Proof of Theorem \ref{cons}:} Sharing the notation in  the proof of Theorem \ref{thm_2_1}, we can write
\[
\frac{{\cal{T}}_{n,w}}{n}=\frac{1}{n}U_{1,n}+\frac{n-1}{n}U_{2,n},
\]
where
\begin{eqnarray*}
U_{1,n} & = & \frac{1}{n} \sum_{j,k=1}^n h(X_j,X_j; \widehat c_n, \widehat \varphi_n)\\
U_{2,n} & = & \frac{1}{n(n-1)} \sum_{1\leq j \neq k \leq n} h(X_j,X_k; \widehat c_n, \widehat \varphi_n)
\end{eqnarray*}
 As $|h(X_j,X_k; \widehat c_n, \widehat \varphi_n)| \leq 4$, it readily follows that $(1/n)U_{1,n} \to 0$ a.s.. To prove the result we show that $U_{2,n}  \stk \Delta_w$. With this aim, we will show  that
 \begin{equation} \label{aux21}
\lim_{d\to 0} \mathsf E\left[ \sup_{\|(c,\varphi)-(c_X, \varphi_X )\|\leq d}\left | 
h(X_1,X_2; c, \varphi)- h(X_1,X_2; c_X, \varphi_X)\right | \right] \to 0.
\end{equation} 
 Then, applying Theorem 2.9 of \cite{Iverson}, the desired result follows.

Notice that $h(x,y; c,\varphi)=h_2(x,y; c,\varphi) -h_1(x; c,\varphi) -h_1(y; c,\varphi)+h_0$, with (assuming that $x \leq y$)
 \begin{eqnarray*}
h_2(x,y; c,\varphi) & = & \int_0^\infty\min \{ 1,t(x/c)^\varphi \} \min \{ 1,t(y/c)^\varphi \}w(y) \: dt\\
 & = & \left(\frac{x}{c}\right)^\varphi \left(\frac{y}{c}\right)^\varphi \int_0^{(c/y)^\varphi} t^2w(t)\: dt+
 \left(\frac{x}{c}\right)^\varphi  \int_{(c/y)^\varphi}^{(c/x)^\varphi} tw(t)\: dt+\int_{(c/x)^\varphi}^{\infty} w(t)\: dt\\
 &:=& h_{21}(x,y; c,\varphi) +  h_{22}(x,y; c,\varphi)+  h_{23}(x,y; c,\varphi),\\
 h_1(x; c,\varphi) & = & \int_0^\infty \min \{ 1,t(x/c)^\varphi \} \psi_0(t) w(y) \: dt\\
 & = & \left(\frac{x}{c}\right)^\varphi  \int_0^{(c/x)^\varphi} t\psi_0(t) w(t) \: dt+ \int_{(c/x)^\varphi}^{\infty} \psi_0(t) w(t)\: dt\\
 &:=& h_{11}(x; c,\varphi) +  h_{12}(x; c,\varphi),\\
  h_0 & = & \int_0^\infty \psi_0(t)^2 w(y) \: dt.
 \end{eqnarray*}
 We have that
  \begin{eqnarray*}
\frac{\partial}{\partial c}  h_{2}(x,y; c,\varphi) & = & -\frac{2\varphi}{c}h_{21}(x,y; c,\varphi)-\frac{\varphi}{c}h_{22}(x,y; c,\varphi),\\
\frac{\partial}{\partial c}  h_{1}(x; c,\varphi) & = & -\frac{\varphi}{c}  h_{11}(x; c,\varphi), \\
\frac{\partial}{\partial \varphi}  h_{2}(x,y; c,\varphi) & = & \log\left(\frac{xy}{c} \right) h_{21}(x,y; c,\varphi)+\log\left(\frac{x}{c} \right) h_{22}(x,y; c,\varphi),\\
\frac{\partial}{\partial \varphi}  h_{1}(x; c,\varphi) & = &  \log\left(\frac{x}{c} \right)  h_{11}(x; c,\varphi).  
\end{eqnarray*}
For each fixed $x,y>0$, taking into account that $0 \leq h_{21}, h_{22}, h_{11}, h_{12} \leq 1$, it readily follows that the above derivatives are bounded when $\|(c,\varphi)-(c_X, \varphi_X )\|\leq d$. Therefore, applying the mean value theorem and recalling that $\mathsf E( |\log(X)|)<\infty$, one gets
\[ \mathsf E\left[ \sup_{\|(c,\varphi)-(c_X, \varphi_X )\|\leq d}\left | 
h(X_1,X_2; c, \varphi)- h(X_1,X_2; c_X, \varphi_X)\right | \right] \leq M d,\]
for some positive constant $M>0$, and \eqref{aux21} holds
\bewend

\bibliographystyle{apalike}


\section*{Funding information}

The work of B. Milo\v sevi\'c is  supported by the Ministry of Science, Technological Development and Innovations of the Republic of Serbia (the contract 451-03-47/2023-01/ 200104), and  by the COST action
CA21163 - Text, functional and other high-dimensional data in econometrics: New models, methods, applications (HiTEc).
M.D. Jim\'enez-Gamero  is supported by grant PID2020-118101GB-I00,
Ministerio de Ciencia e Innovaci\'on (MCIN/ AEI /10.13039/501100011033).

\end{document}